\documentclass[aps,prl,twocolumn,groupedaddress]{revtex4}
\usepackage{amsmath,amssymb,xspace,graphicx}
\newcommand{\taus}{\ensuremath{\tau_\text{S}}\xspace}
\newcommand{\taum}{\ensuremath{\tau_\text{M}}\xspace}
\newcommand{\tauf}{\ensuremath{\tau_\text{F}}\xspace}

\newcommand{\M}{\ensuremath{\mathcal M}\xspace}
\newcommand{\avstress}{\ensuremath{\langle\sigma\rangle}\xspace}
\begin{document}


\title{Instability and spatiotemporal rheochaos in a shear-thickening fluid model}


\author{A.~Aradian}
\email{A.Aradian@ed.ac.uk}
\author{M.E.~Cates}
\affiliation{School of Physics, University of Edinburgh, JCMB
Kings Buildings, Edinburgh EH9 3JZ, United Kingdom.}


\date{\today}

\begin{abstract}
We model a shear-thickening fluid that combines a tendency to form
inhomogeneous, shear-banded flows with a slow relaxational
dynamics for fluid microstructure. The interplay between these
factors gives rich dynamics, with periodic regimes (oscillating
bands, travelling bands, and more complex oscillations) and
spatiotemporal rheochaos. These phenomena, arising from
constitutive nonlinearity not inertia, can occur even when the
steady-state flow curve is monotonic. Our model also shows
rheochaos in a low-dimensional truncation where sharply defined
shear bands cannot form.\end{abstract}


\maketitle

Complex fluids exhibit much interesting behavior under shear, due
to strong couplings between mesoscopic structure and flow.
Experiments show cases where, under steady external driving, an
unstable flow arises, giving a time-dependent strain rate at
constant imposed shear stress, or vice versa. Sustained temporal
oscillations are seen in surfactant mesophases and
solutions~\cite{Wheeler,Pine,Wunenburger,Salmon,Manneville,CourbinPanizza}
and polymer solutions~\cite{HilliouVlassopoulos}, while erratic
temporal responses have been found both in these and in related
materials, e.g., wormlike
micelles~\cite{BandyopadhyayThin,BandyopadhyayThick,Callaghan},
lamellar phases~\cite{Salmon,Salmon2,Manneville} and
colloids~\cite{Laun, Hebraud}. There are strong
indications~\cite{BandyopadhyayThin,BandyopadhyayThick,Salmon}
that these erratic signals result from a deterministic chaotic
dynamics. Chaotic behavior of bulk flows at virtually zero
Reynolds number (negligible inertia) must stem from nonlinearity
within the rheological constitutive equation, and has been dubbed
`rheochaos' \cite{CHA,Chakrabarti,Suzanneprl}.

Such flow instabilities affect both
shear-thinning~\cite{BandyopadhyayThin} and shear-thickening
micellar materials~\cite{BandyopadhyayThick}. Shear-thick\-ening
is, in itself, a widely observed but poorly understood phenomenon
which affects not only micelles but, e.g., dense colloids, where
there is again evidence of bulk rheological
instability~\cite{Laun,Hebraud}. Below we study a simple model for
a shear-thickening fluid at steady controlled stress, which
generalises to spatially inhomogeneous flows a model first
proposed in Ref.~\cite{CHA}. The latter was shown to give simple
oscillations but not chaos (unless an unconvincing `double memory'
term was used). We find that the interplay of a very simple
structural memory with constitutive nonlinearity can, as hoped by
the authors of~\cite{CHA}, give complex dynamics and rheochaos~--
but only if spatial heterogeneity is allowed for. Our work is
related to, but different from, that of Fielding and
Olmsted~\cite{Suzanneprl} which addresses shear-thinning
fluids~\cite{note1}. Also related is Ref.~\cite{Chakrabarti} which
concerns nematic liquid crystals (which can be chaotic even
without spatial heterogeneity). We show below that there are
robust generic features in the rheochaos produced in these various
models, which generally involve a failed attempt to create a
steady shear-banded flow. (In the shear thinning case the bands
have a common stress but unequal strain rates; for us, the reverse
is true.) One distinctive finding of our work is rheochaos in a
model where the steady state flow curve is strictly monotonic.
Here (in contrast with~\cite{Suzanneprl}) no steady banded
solution, stable or otherwise, exists. Yet an innate tendency to
form bands transiently, combined with structural memory that
frustrates their long-term survival, can lead to chaos. A second
unexpected finding is that a low dimensional truncation of our
model also gives chaos, even though sharp interfaces between shear
bands are then suppressed. Thus high-order Fourier components of
spatial variation are not essential to rheochaos.

Candidates for a slow structural mode include the mean length of
wormlike micelles, local composition variables (e.g., colloidal
volume fraction) and `fluidity' parameters~\cite{sendai,fluidity}
reflecting local microstructure, bonding state, etc.. In all cases,
the time scale $\taus$ is distinct from the Maxwell time $\taum$
for linear stress relaxation; and in our work, we assume
$\taus/\taum \geq 1$ (in contrast to~\cite{Suzanneprl}). An
involvement of slowly evolving fluid structure has been clearly
evidenced in some experimental
cases~\cite{Wunenburger,CourbinPanizza,Salmon}.

\paragraph{The model.}
We assume that the shear stress $\sigma$ decouples from other
stress components~\cite{leshouches} and depends only on the rate
of shear strain $\dot\gamma$. For a cylindrical Couette geometry
with axial coordinate $z$, we consider one-dimensional
inhomogeneity along this direction only (as would arise for steady
shear bands in a thickening material~\cite{Olmsted}). We assume
the shear flow is homogeneous within each slice of height $z$,
appropriate to a low-Reynolds limit; $\dot\gamma$ is then uniform
and fixed by the imposed wall velocity. The shear stress
$\sigma(z,t)$ evolves as follows, where $t$ is time and units are
such that the transient elastic modulus is unity:
\begin{equation}
\label{spaceCHA} \dot\sigma(z,t)= \dot\gamma -R(\sigma) - \lambda
\int_{-\infty}^t\hspace{-.3cm} \M(t-t') \,\sigma(z,t')\,
\mathrm{d}t' +\kappa\nabla^2\sigma
\end{equation}
The term $R(\sigma)=a\sigma-b\sigma^2+c\sigma^3$
corresponds to instantaneous nonlinear stress relaxation \cite{CHA}, with $a=
1/\taum$, and $b$, $c$ chosen so that $\sigma(R)$ is an S-shape
(see Fig.~1). This creates a tendency for the fluid to form shear
bands stacked in the $z$-direction. The integral term represents a
retarded relaxation of stress and/or a slow evolution of fluid
structure on timescales $\taus$ (see~\cite{CHA}). $\M$ is a
decaying memory kernel, chosen as $\M(t)=\taus^{-1}
\exp(-t/\taus)$; $\lambda > 0$ governs its strength. Finally, the
nonlocal term, assigning to stress a diffusivity
$\kappa$, selects a unique banded flow in steady state
\cite{LuOlmsted}. Note that our model is closely analogous to the
FitzHugh-Nagumo for neuronal activity~\cite{neuronrefs}, albeit
with an unusual long-range coupling.

\begin{figure}
  \includegraphics[width=.46\textwidth]{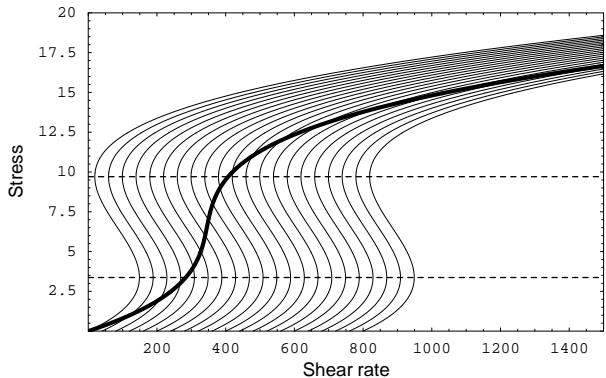}
  \caption{Long-term flow curve $R(\sigma) = \dot\gamma-\lambda \sigma$ (thick line),
and underlying short-term curves obeying $R(\sigma) =
\dot\gamma-\lambda m$ (thin lines, with, from left to right,
$m=0,1,2,\ldots,20$). The stress range between the dotted lines
corresponds to the unstable region (for $\taus=100$). Parameters:
$\lambda=40$, $a=1/\taum=100$, $b=20$, $c=1.02$, $\kappa=0.01$,
$H=1$.}
\end{figure}
\paragraph{Qualitative features.}
We now rewrite equation~\eqref{spaceCHA} as an exactly
equivalent differential system:
\begin{equation}
\label{spaceCHAdifferential}
\dot\sigma=\dot\gamma-R(\sigma)-\lambda
m+\kappa\nabla^2\sigma\quad\text{and}\quad\dot
m=-\frac{m-\sigma}{\taus}
\end{equation}
The integral term in eq.~\eqref{spaceCHA} has become an auxiliary
variable $m(z,t)$, which we call the memory. The memory encodes
the delayed (structural) part of the stress relaxation, and, for
our choice of $\M$, decays with rate $\taus^{-1}$ towards the
local value of the stress $\sigma(z,t)$~\cite{note2}. The homogeneous
version of the present model (no
$z$-dependence) shows temporal instability, leading to van der Pol
type oscillations~\cite{CHA}.

The steady-state flow curve $\sigma(\dot\gamma)$ for the model is
obtained from $\dot\sigma=\dot m=\nabla^2\sigma =0$, as
$R(\sigma)+\lambda\sigma = \dot\gamma$. At each point on the
curve, the memory $m$ has relaxed to the steady state stress
($m=\sigma$). However, this relaxation involves the long timescale
$\taus$. At times much shorter than $\taus$, the fluid will
instead behave as if the memory $m$ were frozen: one has a set of
`instantaneous' flow curves $R(\sigma)+\lambda m = \dot\gamma$.
Thus, despite the fact the steady-state flow curve is monotonic,
the existence of fixed $m$
curves with an S-shape gives rise, on short times, to spatial
inhomogeneity in the form of shear bands along $z$ (Fig.~1). As
these bands are subjected to the local van der Pol instability
referred to above, a very rich spatio-temporal dynamics arises as
we now show.
\begin{figure}
  \includegraphics[width=.44\textwidth,clip=true]{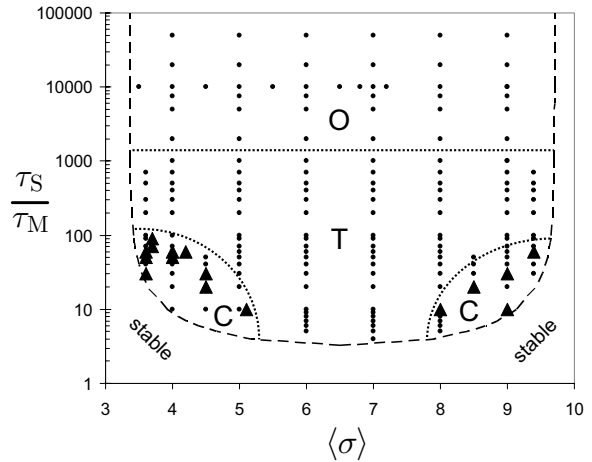}
  \caption{Phase diagram of the model when $\taus$ and $\avstress$
  are varied: ($\blacktriangle$)~chaotic point, ($\bullet$)~periodic point.
  Three main regimes are observed: (O)~oscillating shear bands,
  (T)~travelling shear bands, (C)~chaotic regions. The outer dashed line
  is the linear stability limit,  $R'(\avstress)+1/\taus=0$~\cite{CHA,Aradian}. The dotted lines
between regions are guides to the eye; there are no sharp transitions.
The C-regions enclose all observed chaotic
points, but contain internal structure with
periodic and chaotic pockets. Numerical parameters
as in Fig.~1.}
\end{figure}
\paragraph{Fourier-Galerkin truncation.}
Equations~\eqref{spaceCHAdifferential} were solved numerically
using a spectral Galerkin truncation~\cite{Boyd}, with
$\sigma(z,t)$ and $m(z,t)$ decomposed in Fourier modes
$\sigma_k(t)$ and $m_k(t)$:
$\sigma(z,t)=\sum_{k=0}^{N-1}\sigma_k(t) \cos(k\pi z/H)$ etc.,
with $H$ the axial extent of our Couette device ($0\leq z \leq H$)
and $N$ the order of the truncation. (Sine modes are excluded by
zero stress-flux boundary conditions, $\nabla\sigma=0$.) The
truncation thus replaces eqs.~\eqref{spaceCHAdifferential} for
$\sigma(z,t)$ and $m(z,t)$ by $2N$ coupled ordinary differential
equations for the modes $\sigma_k(t),m_k(t)$. The results
presented below used a high-resolution truncation ($N=40$);
smaller $N (\geq 20)$ already give similar results. We work at
fixed torque on the Couette, that is, at fixed value of the
spatially averaged stress $\avstress=\int_0^H\sigma(z,t)dz/H$
\cite{Aradian}.
\paragraph{Phase diagram.}
In Fig.~2 we show a `phase diagram' grouping the different types
of spatiotemporal dynamics seen \cite{warning}
on varying $\avstress$ and
$\taus/\taum$ (with $\taum=0.01$, and remaining parameters held
at the values of Fig.~1, and small amplitude noise in the
initial condition). Remarkably,
despite the high-dimensional dynamical system under
consideration ($2N=80$), the phase diagram displays a simple
overall structure, with three main regimes: periodic
response with oscillating shear bands at extremely long \taus;
periodic response with travelling bands at long \taus; and chaotic
response at shorter \taus and off-centered values of \avstress.
\begin{figure}
  \includegraphics[width=.49\textwidth,clip=true]{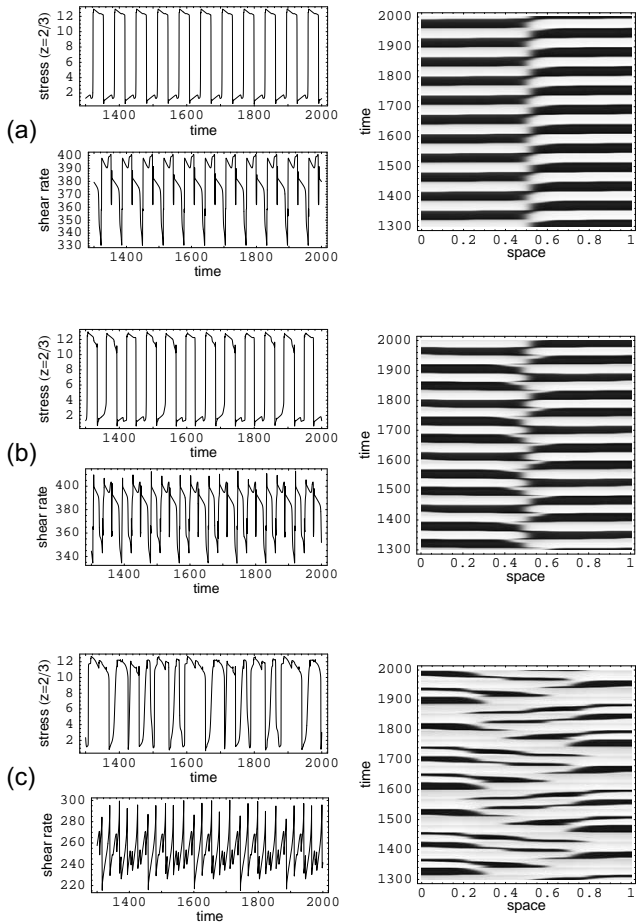}
  \caption{Typical responses in the oscillating shear band regime, chosen along the line
  $\taus/\taum=10^4$ in the phase diagram. (a)~Flip-flopping bands at $\avstress=7$;
   (b)~Zig-zagging interface at $\avstress=7.1$; (c)~Complex periodic motion at $\avstress=9$.
   Each group presents time series of the stress $\sigma$ at $z=2/3$, the shear rate $\dot\gamma$,
   and a space-time plot of $\sigma(z,t)$ with $t$ vertical, $z$ horizontal (clear shades correspond
   to high stress, dark shades to low stress).}
\end{figure}
%
\paragraph{Oscillating shear bands.}
The succession of behaviors observed along the horizontal line
$\taus/\taum=10^4$ in the phase diagram typifies this regime.
Near the middle of the line (e.g.
$\avstress=7.0$) we find two `flip-flopping' bands: as
Fig~3-a shows, mid-cycle a low-shear band and a
high-shear band each span about half of the cell. Each undergoes a
local van der Pol type
oscillation; to keep $\langle\sigma\rangle$ fixed as prescribed,
these are synchronous and each half-cycle the identity of the
two bands is reversed.
The time series of the local stress
(e.g., $\sigma(2/3,t)$; Fig~3-a) shows a
large amplitude oscillation, close to a square wave, with a
flip-flop period $\tauf$ of order \taus. The shear rate
$\dot\gamma$ (independent of $z$) accordingly shows a periodic
evolution, with a more complicated waveform.

If we now raise or lower $\langle\sigma\rangle$ slightly (e.g.,
$\avstress=7.1$, Fig.~3-b), in addition to the bands'
flip-flopping, their interface adopts a zig-zag motion.
(Imposition of $\langle\sigma\rangle$ now enforces unbalanced
proportions of the low- and high-shear bands; hence on flipping,
the interface {\em must} move to and fro to maintain these
proportions throughout the cycle.) The time series $\sigma(2/3,t)$
is slightly distorted, with period now a multiple (here three) of
$\tauf$. Moving further towards the wings of the phase diagram,
the intrinsic flipping dynamics of the bands, coupled to the
global constraint on $\langle\sigma\rangle$ (e.g., $\avstress =
9.0$, Fig.~3-c) gives an extremely complex motion which,
remarkably, manages to remain periodic. The time series
$\sigma(2/3,t)$ becomes very ragged and the period is a large
multiple of $\tauf$.
\paragraph{Travelling shear bands.} For shorter
\taus ($20\le \taus/\taum \le 10^3$) we find a new periodic regime
in which shear bands nucleate at a boundary and then cross the
system, usually one at a time, at roughly constant velocity
(Fig.4). Rarely, bands can nucleate periodically from a
non-boundary point~\cite{Aradian}, giving two outgoing bands in a
one-dimensional analogue of the `target patterns' seen in chemical
oscillators~\cite{TargetPatterns}. Details here depend strongly on
the initial noise.
\begin{figure}
  \includegraphics[width=.48\textwidth,clip=true]{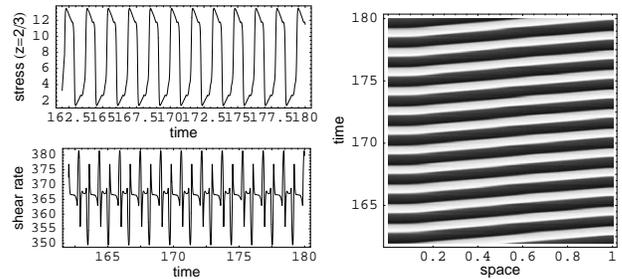}
  \caption{Time series and space-time plot of the stress $\sigma(z,t)$
  in the travelling band regime. Parameters: $\taus/\taum=90$,
  $\avstress=7$, others as in Fig.~1.}
\end{figure}
\begin{figure}
  \includegraphics[width=.48\textwidth,clip=true]{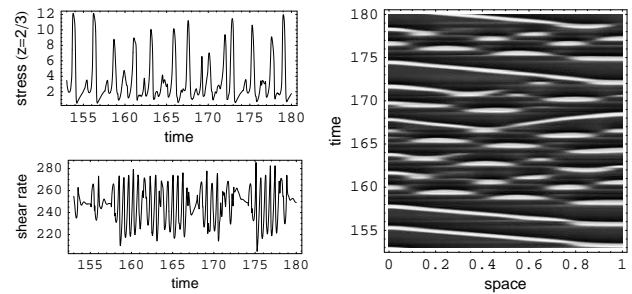}
  \caption{Time series and space-time plot of the stress $\sigma(z,t)$ in the chaotic
  bands regime (`bubbly' type). Parameters: $\taus/\taum=90,\avstress=3.6$, others as in Fig.~1.}
\end{figure}
The travelling bands are kinematic waves arising from staggered
phase distribution of the local van der Pol-like oscillation. Our
data~\cite{Aradian} also indicate that the band velocity decreases
as \taus is increased. Somewhat similar band motion was reported
in Ref.\cite{Suzanneprl} but there the bands `ricochet' off the
walls of the container; ours disappear and renucleate at the
opposite wall.
%
%
\paragraph{Spatiotemporal rheochaos.}
The third regime is chaotic and arises mainly for
$10\le\taus/\taum\le 100$, in the wings of the phase diagram. A typical
dynamics in this regime (Fig.~5) comprises a `bubbly' phase of
localized, short-lived shear bands that appear and disappear
erratically in the cell. (Occasionally a band survives longer and
shoots across the cell.) Other types of chaotic space-time
patterns were also found~\cite{Aradian}: `wiggling'
travelling bands and `defect dominated' regimes (the latter
close to those found for shear thinning micelles
Ref.~\cite{Suzanneprl}). Chaos, for each point marked in
Fig.~2, was confirmed by a positive Lyapunov
exponent~\cite{Aradian}.

\paragraph{Role of stress diffusion.} The
diffusivity $\kappa$ has a singular role in steady-state shear
banding, selecting a unique banded flow~\cite{LuOlmsted}. However
in our work stress diffusion is, in Fourier space, a diagonal term
acting mainly by damping the higher modes: small $\kappa$
increases the order $N$ required for realistic truncation, and
this tends to favor chaos. We suspect that $\kappa$ also plays a
role in the crossover from travelling to oscillating bands seen at
$\taus/\taum\simeq 10^3$ in Fig.~2: for our parameters, this
happens roughly when $\taus$ becomes equal to the typical
diffusion time across the Couette, $\tau_\text{diff}=1/\kappa q^2$
(with $q=\pi/H$). We do find the crossover to move towards longer
timescales as $\kappa$ is reduced~\cite{Aradian}.
%
%
\paragraph{Low-truncation limit; route to chaos.}
Our work, with that of \cite{Suzanneprl,Chakrabarti}
suggests a generic interpretation of rheochaos in terms of
the erratic motion of discrete band interfaces~\cite{CHA}.
But if sharp interfaces were essential, there would be no chaos in
a low-mode truncation that allows only smooth variation of
$\sigma,m$ on the scale of the system. To test this, we
studied the case $N=3$; with $\sigma_0$ and $m_0$ fixed by
$\avstress$, our dynamical variables comprise $\sigma_1(t)$,
$\sigma_2(t)$, $m_1(t)$ and $m_2(t)$. Our numerics
show~\cite{Aradian} that, although shrunk, the chaotic regions persist.
Sharply resolved band interfaces are thus not essential to
rheochaos.
Unlike the (numerically more delicate)
high-resolution truncation, the $N=3$ case allows
determination of the transition to
chaos as \avstress is varied at fixed \taus: we found
a classical period-doubling cascade.
\paragraph{Concluding remarks.}
Our minimal model of a shear-thickening fluid combines the
tendency to form shear bands with structural
relaxation. This combination leads to
oscillatory and travelling banded solutions and to
spatiotemporal chaos. Our work strongly supports a
generic `frustrated shear-banding' picture of rheochaos (e.g.
\cite{CHA}) as instanced in recent work on two other models
\cite{Suzanneprl, Chakrabarti}:
it seems increasingly clear that spatial inhomogeneity plays a key
role in rheochaos. Intriguingly, in our model it is enough to
represent this by a very low dimensional truncation of the spatial
structure: transient shear bands, in the sense of sharp interfaces
between layers of material of different flow history, are
\emph{not} essential. But, in practice, the stress diffusivity
$\kappa$ usually is small and ensures rather sharp interfaces; we
suspect the `frustrated banding' picture to hold in many of
the experimental realizations of rheochaos so far reported.

We thank L.~B{\'e}cu, A.~Colin,
S.~Fielding, S.~Manneville, P.~Olmsted, D.~Roux and J.-B.~Salmon
for discussions. AA funded by EPSRC grant GR/R95098.
%
%
%
%
%

%
%
%

\begin{thebibliography}{1}
%
%
%
\bibitem{Wheeler}
E.~K.~Wheeler, P.~Fischer and G.~G.~Fuller, \emph{J.~Non-Newt.\
Fluid Mech.} {\bf 75}, 193 (1998).
%
\bibitem{Pine}
Y.~T.~Hu, P.~Boltenhagen, E.~Matthys, D.~J.~Pine,
\emph{J.~Rheology} {\bf 42}, 1209 (1998).
%
\bibitem{Wunenburger}
A.-S.~Wunenburger {\it et al.},
\emph{Phys.\ Rev.\ Lett.} {\bf 86}, 1374 (2001).
%
\bibitem{Salmon}
J.-B.~Salmon, A.~Colin and D.~Roux \emph{Phys.\ Rev.\ E} {\bf 66},
031505 (2002).
%
\bibitem{Manneville}
S.~Manneville, J.-B.~Salmon and A.~Colin, \emph{Eur.\ Phys.\ J.~E}
{\bf 13}, 197 (2004).
%
\bibitem{CourbinPanizza}
L.~Courbin, P.~Panizza and J.-B.~Salmon, \emph{Phys.\ Rev.\ Lett.}
{\bf 92}, 018305 (2004).
%
\bibitem{HilliouVlassopoulos}
L.~Hilliou and D.~Vlassopoulos, \emph{Ind.\ Eng.\ Chem.\ Res.\ }
{\bf 41}, 6246 (2002).
%
\bibitem{BandyopadhyayThin}
R.~Bandyopadhyay, G.~Basappa and A.~K.~Sood, \emph{Phys.\ Rev.\
Lett.\ }{\bf 84}, 2022 (2000).
%
\bibitem{BandyopadhyayThick}
R.~Bandyopadhyay and A.~K.~Sood, \emph{Europhys.\ Lett.\ }{\bf 56}
447 (2001).
%
\bibitem{Callaghan}
W.~M.~Holmes, M.~R.~L{\'o}pez-Gonz{\'a}lez and P.~T.~Callaghan,
\emph{Europhys.\ Lett.\ } {\bf 64}, 274 (2003).
%
\bibitem{Salmon2}
J.-B.~Salmon, S.~Manneville and A.~Colin, \emph{Phys.\ Rev.~E}
{\bf 68}, 051504 (2003).
%
\bibitem{Laun}
Laun, H.~M., \emph{J.\ Non-Newt.\ Fluid Mech.}{\bf\ 54}, 87--108
(1994).
%
\bibitem{Hebraud}
D.~Lootens, H.~Van Damme and P.~H{\'e}braud, \emph{Phys.\ Rev.\
Lett.\ } {\bf 90}, 178301 (2003).
%
\bibitem{CHA}
M.~E.~Cates, D.~A.~Head and A.~Ajdari, \emph{Phys.\ Rev.~E} {\bf
66}, 025202 (2002).
%
\bibitem{Chakrabarti}
B.~Chakrabarti {\it et al.},
\emph{Phys.\ Rev.\ Lett.\ } {\bf 92}, 055501 (2004).
%
\bibitem{Suzanneprl}
S.~M.~Fielding and P.~D.~Olmsted, \emph{Phys.\ Rev.\ Lett.\ } {\bf
92}, 084502 (2004).
%
\bibitem{note1} Shear thickening and thinning differ fundamentally:
one cannot simply interchange stress and strain-rate variables within
a dynamical model.
%
\bibitem{fluidity}
C.~Derec, A.~Ajdari and F.~Lequeux, \emph{Eur.\ Phys.~J.~E} {\bf
4}, 355 (2001).
%
\bibitem{sendai} A.~Aradian and M.~E.~Cates, in \emph{Proc. of
the 3rd Int.\ Symp.\ on Slow Dynamics in Complex Systems}, AIP
Conference Proceedings vol. 708, M.~Tokuyama and I.~Oppenheim,
Eds. AIP, Melville (NY), 2004.
%
%
\bibitem{leshouches}
M.~E.~Cates, in \emph{Slow relaxations and nonequilibrium dynamics
in condensed matter, Les Houches Session LXXVII}, J.-L.~Barrat
\emph{et al.}, Eds. Springer, New York, 2003.
%
\bibitem{Olmsted}
J.~L.~Goveas and P.~D.~Olmsted, \emph{Eur.\ Phys.\ J.~E} {\bf 6},
79 (2001).
%
\bibitem{LuOlmsted}
C.-Y.~D.~Lu, P.~D.~Olmsted and R.~C.~Ball, \emph{Phys.\ Rev.\
Lett.\ } {\bf 84}, 642 (2000).
%
\bibitem{neuronrefs}
J.~Keener and J.~Sneyd, \emph{Mathematical Physiology}. Springer,
New York, 1998.
%
\bibitem{note2}
The form of $\M$ should not be important~\cite{CHA}; our
exponential choice admits accurate and efficient numerics.
%
\bibitem{Boyd}
J.~P.~Boyd, \emph{Chebyshev and Fourier Spectral Methods}, Dover
Publications, New York, 2000.
%
\bibitem{Aradian}
A.~Aradian and M.~E.~Cates, in preparation.

\bibitem{warning} As is expected in
chaotic dynamics, there is some dependence on the numerical
schemes used; see~\cite{Aradian}.
%
\bibitem{TargetPatterns}
I.~R.~Epstein and J.~A.~Pojman, \emph{An Introduction to Nonlinear
Chemical Dynamics}. Oxford University Press, Oxford, 1998.
%
\end{thebibliography}
\end{document}